\pdfoutput=1
\documentclass[letterpaper,superscriptaddress,aps,pra,nolongbibliography,twocolumn,showpacs,floatfix,10pt]{revtex4-2} 
\usepackage{url}
\usepackage{amsmath}
\usepackage{amssymb}
\usepackage[utf8]{inputenc}
\usepackage[english]{babel}
\usepackage[T1]{fontenc}

\usepackage{tikz}
\usetikzlibrary{arrows}

\usepackage{amsmath}
\usepackage{hyperref}
\usepackage{lipsum}
\usepackage{graphicx,helvet}
\usepackage{color}
\usepackage{mathtools}
\usepackage{bbm,bm}
\usepackage{soul}
\usepackage{amsfonts}
 \usepackage{amsmath}
\usepackage{lipsum}
\definecolor{mygray}{gray}{0.4}
\definecolor{light-blue}{rgb}{0.8,0.85,1}
\graphicspath{{images/}}
\usepackage{amsthm}
\mathchardef\Re="023C
\mathchardef\Im="023D

\newcommand{\jami}{Jamiołkowski}

\newcommand{\mcL}{\mathcal{L}}
\newcommand{\mcB}{\mathcal{B}}

\newcommand{\mcE}{\ensuremath{\mathcal{E}}}
\newcommand{\mcG}{\ensuremath{\mathcal{G}}}

\newcommand{\nnn}{\mathcal{N}}
\newcommand{\choi}{\ensuremath{\mcD}}
\newcommand{\mmm}{\mathcal{M}}
\newcommand{\sss}{\mathcal{S}}
\newcommand{\mcD}{\mathcal{D}}

\newcommand{\valpha}{{\vec \alpha}}
\newcommand{\vgamma}{{\vec \gamma}}
\newcommand{\one}{\openone}

\newcommand{\hilbert}{\ensuremath{{\sf H}}}

\newcommand{\id}{\text{id}}
\newcommand{\pce}{\ensuremath{\text{PCE}}}
\newcommand{\vbeta}{\vec \beta}

\newcommand{\ie}{i.e.}

\newcommand{\eref}[1]{Eq.~(\ref{#1})} 
\newcommand{\sref}[1]{sec.~\ref{#1}}
\newcommand{\fref}[1]{Fig.~\ref{#1}}

\newcommand{\Eref}[1]{Eq.~(\ref{#1})} 

\newcommand{\Fref}[1]{Fig.~\ref{#1}}

\newcommand{\tr}{\mathop{\mathrm{Tr}}u\nolimits}

\newcommand{\ket}[1]{{\vert #1 \rangle}}
\newcommand{\bra}[1]{{\langle #1 \vert}}
\newcommand{\proj}[2]{{\vert #1 \rangle \langle #2 \vert}}
\newcommand{\projj}[1]{{\vert #1 \rangle \langle #1 \vert}}

\usepackage{physics}
\usepackage{breqn}
\newcommand{\paulicomponents}{r_\valpha}
\newcommand{\taus}{\tau_\valpha}
\newcommand{\pceg}{\mcG_{\vec \alpha}}
\newcommand{\appref}[1]{appendix~\ref{#1}}

\def\kket#1{\mathinner{|{#1}\rangle \! \rangle}}
\def\ddyada#1{\mathinner{|{#1}\rangle\!\rangle\!\langle\!\langle{#1}|}}
\def\ddyad#1#2{\mathinner{|{#1}\rangle\!\rangle\!\langle\!\langle{#2}|}}
\def\bbrakket#1#2{\mathinner{\langle\!\langle {#1}|{#2}\rangle\!\rangle}}

\newcommand{\unam}{Universidad Nacional Aut\'onoma de M\'exico, Ciudad de M\'exico 01000, Mexico}
\newcommand{\icf}{Instituto de Ciencias F\'{\i}sicas, Universidad Nacional Aut\'onoma de M\'exico, Cuernavaca 62210, Mexico}
\newcommand{\ifunam}{Instituto de F\'{\i}sica, \unam}

\newcommand{\sas}{Institute of Physics, Slovak Academy of Sciences, D\'ubravsk\'a cesta 9, Bratislava 84511, Slovakia}

\newcommand{\ecfmUsac}{Instituto de Investigaci\'on en Ciencias F\'isicas y Matem\'aticas, Universidad de San Carlos de Guatemala, Ciudad Universitaria, Guatemala 01012, Guatemala}
\newcommand{\affalejandro}{Departamento de Física, CCEN, Universidade Federal de Pernambuco, Recife 50670-901, PE, Brazil}
\begin{document}
\title{Pauli component erasing quantum channels} 
\author{José Alfredo de León} \affiliation{\ecfmUsac}\affiliation{\ifunam}
\author{Alejandro Fonseca} \affiliation{\affalejandro}\affiliation{\ifunam}
\author{François Leyvraz} \affiliation{\icf}
\author{David Davalos} \affiliation{\sas}
\author{Carlos Pineda} \email{carlospgmat03@gmail.com} \affiliation{\ifunam}
\begin{abstract} 
Decoherence of quantum systems is described by quantum channels.  However, a
complete understanding of such channels, especially in the multi-particle
setting,  is still an ongoing difficult task.  We propose the family of quantum
maps that preserve or completely erase the components of a multi-qubit system
in the basis of Pauli strings, which we call {\it Pauli component erasing} 
maps.  For the corresponding channels, it is shown that the preserved
components can be interpreted as a finite vector subspace, from which we derive
several properties and complete the characterization.  Moreover, we show that
the obtained family of channels forms a semigroup and derive its generators. We
use this simple structure to determine physical implementations and connect the
obtained family of channels with Markovian processes.
%
\end{abstract} 
 
\maketitle
\newcommand{\sa}{\ensuremath{\textit{a}}}
\newcommand{\san}{\ensuremath{\textit{A}}}
\section{Introduction} 

Quantum correlations~\cite{PhysicsPhysiqueFizika.1.195,
PhysRevLett.126.170404,Ollivier2001}, including entanglement~\cite{Horodecki},
are an important resource for a wide variety of tasks that
include teleportation~\cite{PhysRevLett.70.1895}, quantum
computation~\cite{nielsen_chuang_2011}, and others~\cite{Sapienza2019}.
However, this resource is also extremely
delicate~\cite{nielsen_chuang_2011,Schlosshauer2004}, especially 
for multi-particle systems~\cite{Horodecki}; that is why an important part
of the efforts of the community implementing quantum technologies is 
devoted to tackle this issue from an experimental \cite{Mooney_2021,Briegel1998} and
theoretical \cite{Bennett1996,Georgescu2020,RevModPhys.87.307} point of view. 
The process by which quantum correlations are unintentionally dissipated
is called {\it decoherence}~\cite{Schlosshauer2004,sep-qm-decoherence}.
One of the main tools to study the effects of decoherence are 
quantum channels. Quantum channels can describe quantum
noise~\cite{davalosdivisibility,ruskai}, open quantum systems
dynamics~\cite{nielsen_chuang_2011,breuerbook}, and recently even coarse
graining~\cite{Duarte2017,pinedacg}. 
One of the main difficulties 
in characterizing quantum channels is that, like for quantum states, 
the number of parameters required for their description increases quite 
rapidly with Hilbert-space dimension.
%
Moreover, such parameters are constrained in a complicated way by 
physical conditions, such as complete positivity ~\cite{zimansbook}.
%
Describing in detail families of channels having a given property provides
insight into the jungle of quantum operations.
For the qubit case there are several studies concerning the unital case, for which
non-trivial properties can be described using only three parameters, which in
turn form the well-known tetrahedron of Pauli
channels~\cite{Rybar2012,ruskai,davalosdivisibility}. 
More generally, in~Ref.~\cite{nathanson2007pauli} the authors study families of 
convex combinations of quantum-classical channels that relate to
unital qubit channels with positive eigenvalues, and give a
generalization of the Bloch sphere. 
Similarly, a generalization of Pauli channels based on mutually unbiased
measurements is introduced and studied in Ref.~\cite{Siudzinska2020}. Other
studies of channels beyond the qubit can be
found~\cite{pub.1117408526,pub.1060516327,pub.1014660854,Fonseca2019}. 

In this paper we present a generalization of idempotent Pauli channels---\ie{},
the qubit flip operations (bit, phase, and bit-phase when the flip probability is $1/2$), total depolarizing qubit
channel and the identity channel---to the case of $N$ qubits. 
The generalization is done by extending
Pauli observables to \textit{Pauli strings} (tensor products of Pauli
matrices)~\cite{KIMURA2003339,Lawrence2002}. The resulting maps are unital and
diagonal in the Pauli strings' basis. 
We shall in the following refer to 
such maps as {\em Pauli component erasing} (\pce{}) {\it maps}.

The main task which we perform in this paper is the identification of the
conditions which an arbitrary \pce{} map must satisfy in order to be completely
positive. The answer turns out to involve a strikingly simple and unexpected
mathematical structure that is exploited to gain deeper understanding
on aforementioned channels, as we show in section \ref{sec:vector_spaces}. This
structure allows us, for example, to describe such channels with a much reduced
set of parameters (as compared to specifying a list of all erased Pauli
components) or to define an
interesting semigroup structure on the set of all \pce{} channels. 
Additionally, these channels are, in a sense, the simplest possible channels,
and as such can be used as building blocks of more general channels. For 
instance, one can combine them (through convex superposition) or compose them
with unitary transformations.
To summarize succinctly the final result, we show that it is possible
to assign to every Pauli string a simple \pce{} channel, obtained by extending
the system with an ancilla of a single qubit, acting on the combined system by
a unitary involving the Pauli string and tracing over the ancilla. It then
follows from our results that {\em all\/} \pce{} channels arise from such
channels by composition.


%
\par
The paper is organized as follows. In section~\ref{sec:pce_maps} we recall
the properties of quantum channels needed to proceed with the definition 
of 
\pce{} maps. 
In section~\ref{sec:math} we diagonalize analytically the Choi matrix for
arbitrary \pce{} maps and characterize their complete positivity by
interpreting \pce{} quantum channels as finite vector subspaces.
We study the
generators of the semigroup structure associated to the set of \pce{} channels in section~\ref{sec:generators}, and we use
them to derive meaningful physical interpretations of \pce{} channels in
section~\ref{sec:kraus}, as well as Kraus operators of the generators.
To finish, we conclude and discuss future perspectives and possible generalizations in
section~\ref{sec:conclusions}.

\section{Pauli component erasing maps}
\label{sec:pce_maps}

In this section we introduce the family of PCE
maps. Let us start our discussion with a brief review of several basic concepts
of quantum channels that will allow us to introduce some notation, 
and finish with the definition of PCE maps and some generalities. We further 
introduce a useful graphical representation  for them.
\subsection{Quantum channels} \label{subsec:qtm_ch}
Quantum channels are the most general linear operations that a quantum system
undergoes independently of its past~\cite{zimansbook,cirac}.  The physical system
under study will be associated with a Hilbert space denoted by \hilbert{}, and
the set of linear operators over such space will be denoted by
$\mcB(\hilbert)$. That way, a density matrix $\rho$ of such system is an
element of $\mcB(\hilbert)$.

The construction of quantum channels includes basically three ingredients:
linearity, trace preservation, and complete positivity. Linearity is needed to
map every convex combination of density matrices into a convex combination of the
evolution of such density matrices. The
trace preserving property is required for the process $\mcE$ 
 to happen with probability $1$, and reads $\tr\mcE[\rho]=\tr \rho =1$. 
The
complete
positivity condition is needed to preserve positive semidefiniteness and handle
the non-local nature of quantum theory.
A linear map $\mcE$ is positive if it maps density operators to density operators, \ie{}
if $\mcE[\rho]\geq 0$ for all density matrices $\rho$. 
%
%
On the other hand, if one extends a positive map to include an ancilla, the resulting
map is not always positive. If, for an ancillary system of arbitrary dimension, such extension
results in a positive map, we say that the original map is completely
positive~\cite{bengtsson_zyczkowski_2017}.
Quantum channels are required to be completely positive
so as to allow the proper evolution of potentially entangled states with an
ancilla; to test this condition we require some additional steps. 

A simple algorithm to test the complete positivity of a quantum channel 
was developed by 
\jami{}~\cite{jamil} and Choi~\cite{choi}.
One first exploits the isomorphism that maps a channel $\mcE$ to the state
$\choi=\left(\id \otimes \mcE\right)[\proj{\Omega}{\Omega}]$, where
$\ket{\Omega}=1/\dim(\hilbert)\sum_i^{\dim(\hilbert)}\ket{i}\ket{i}$ is a 
maximally entangled state between the original system and an ancilla and
``$\id$'' is the identity channel.
Remarkably, the map \mcE{} is completely positive
if and only if \choi{} (also called the Choi or dynamical matrix of $\mcE$) 
is positive semidefinite~\cite{jamil,choi}. 

\subsection{Structure of PCE maps} 
\label{sec:single:qubit}
%
We have discussed the main features of quantum channels, and now
we turn our attention to introduce the Pauli component erasing maps. We start
by exploring the single-qubit scenario and then we treat the $N$-qubit case.

\begin{figure} 
\begin{center}
\includegraphics{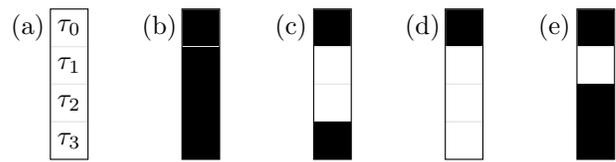}
\end{center}
	\caption{
In (a) we introduce the notation
in the diagrams that represent the single-qubit \pce{} maps,
so that each square corresponds to a single 
$\tau_{\alpha}$, $\alpha=0,1,2,3$. The diagrams in 
(b), (c) and (d) correspond to the identity map, completely dephasing channel,
and 
complete depolarization, respectively, 
as the color of each square indicates the value attained by the corresponding
$\tau_{\alpha}$, either 0 (white) or 1 (black). 
In (e) we show a map that only erases the component
$r_1$, collapsing the Bloch sphere into a disk, and thus does not correspond to
a quantum channel.
}
	\label{fig:one:qubit:examples}
\end{figure}

The most general single-qubit density matrix can be written as
\begin{equation}
\rho=\frac12 {\sum_{\alpha=0}^3 r_\alpha \sigma_\alpha},
\end{equation}
%
with $\sigma_0= \one$, and $\sigma_{1,2,3}$ the usual Pauli matrices.  
Normalization requires that $r_0=1$ and the remaining $r_{1,2,3}$ 
form the Bloch vector. 
Consider the map that projects
each component 
in the following way:
\begin{equation}
r_\alpha \mapsto \tau_\alpha r_\alpha
\label{EqPCE}
\end{equation}
where $\tau_\alpha$ is either $0$ or $1$ (trace preserving requires
that $\tau_0=1$). 
From now on we refer to any operation like that described in
\eref{EqPCE}, as a single-qubit \pce{} map.
Not every such operation  is a quantum channel;
for example, collapsing the entire Bloch ball to a disk on the $xy$
plane ($\tau_{1}=\tau_{2}=1$ and  $\tau_3=0$) 
leads
to a violation of the complete positivity condition. Indeed, a direct evaluation
of such conditions yields~\cite{zimansbook,davalosdivisibility}
%
%
\begin{align}
1+\tau_1+\tau_2+\tau_3 &\geq 0, \nonumber \\
1+\tau_\alpha-\tau_\beta-\tau_\gamma &\geq 0 \ \ \forall ~\alpha\neq \beta \neq \gamma,
\end{align}
where trace preserving is already imposed, and shows that five out of
the eight single-qubit PCE maps are quantum channels.
%
These operations are the identity map, the completely depolarizing channel
($\rho \mapsto \one/2$), as well as the bit, phase, and the bit-phase flip
(with flip probability of $1/2$)
channels~\cite{chuangbook}, 
and can be pictured using one column tables showing the positions of $0$s and
$1$s, see~\fref{fig:one:qubit:examples}.

In order to present and develop the $N$-qubit case, it is useful 
to introduce 
the so-called \textit{Pauli strings}, defined as 
\begin{equation}
\sigma_{\vec \alpha}=\sigma_{\alpha_1}\otimes \sigma_{\alpha_2} \otimes \dots \otimes \sigma_{\alpha_N},
\label{eq:pauli_strings}
\end{equation}
where $\vec \alpha$ denotes a multi-index $\qty(\alpha_1,\ldots,\alpha_N)$
and $\alpha_i=0,1,2,3$. These hermitian operators form an orthogonal 
basis in the space of operators acting on $N$ qubits. In fact, 
$\tr \sigma_{\vec \alpha} \sigma_{\vec \alpha'}=2^N\delta_{\valpha \valpha'}$
and 
$\tr \sigma_{\vec \alpha}= 2^{N}\delta_{\vec\alpha \vec 0}$.
%

Similarly to the single-qubit case, the density matrix $\rho$ of a system of $N$
qubits can be written using Pauli strings in the following way, 
\begin{equation}\label{eq:N_qubits_rho}
	\rho =\frac{1}{2^N} 
            \sum_\valpha r_\valpha \sigma_\valpha,
\end{equation}
so $r_{\vec \alpha}=\langle \sigma_{\vec \alpha} \rangle=\tr \left(\rho
\sigma_{\vec \alpha} \right)$ is the coefficient corresponding to the expansion
of the density matrix in the normalized basis of Pauli strings.  Again,
normalization of the state requires that $r_{\vec 0}=1$.
We shall refer to $\paulicomponents$ as the {\it Pauli components} of the density
matrix of a system of qubits. 

%


In general, a PCE map is a map
that either preserves or completely erases the Pauli components of a
density matrix. That is, 
\begin{equation} \label{eq:PCE_definition}
	\paulicomponents \mapsto \taus \paulicomponents,\quad 
	\taus = 0,1.
\end{equation}
In addition, for the operation to be trace preserving, it is required that $\tau_{\vec
0}=1$. 
It is worth noticing that, as for the single-qubit case, not all PCE maps are
quantum operations. 
On the other hand, constructing and
evaluating the conditions for complete positivity is non-trivial and is the main
problem addressed in this paper.
We shall refer to the map  $\paulicomponents \mapsto \taus \paulicomponents$, 
with arbitrary values of 
$\taus$ (only restricted by complete positivity), as {\it Pauli diagonal maps}.

\begin{figure} 
\includegraphics{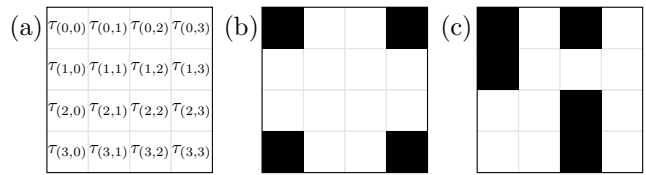}
	\caption{In (a) we introduce the
positions of two qubit diagrams. The diagram in 
(b) corresponds to a quantum channel that results from the tensor product of
bit flip channels in each qubit [see \fref{fig:one:qubit:examples}(c)],
and in (c) a diagram of a map that is not a quantum channel is presented.}
	\label{fig:two:qubit:examples}
\end{figure}

A graphical representation for PCE maps may be introduced, with the two-qubit
case proving to be the most useful.
Consider a $N$-dimensional 
Cartesian grid, with $4^N$ places.  Each place has $N$ integer coordinates, ranging
from 0 to 3, so each place corresponds to a given $\vec \alpha$ in 
\eref{eq:N_qubits_rho}. For a given PCE, we shall fill  
the square if the corresponding $\tau_{\vec \alpha}=1$. Otherwise, we leave it 
empty. 
Examples for $N=1$ and $2$ are provided in \fref{fig:one:qubit:examples} and
\fref{fig:two:qubit:examples}, respectively.

It is worth noticing that the set of \pce{} maps overlaps with the set
of ``Pauli diagonal channels constant on axes'' defined in
Ref.~\cite{nathanson2007pauli}, consisting of convex combinations of
\textit{quantum-classical} channels. In particular, it can be shown that
quantum-classical channels defined with the eigenbasis of some set of $2^N-1$
commuting Pauli observables~\cite{Lawrence2002} comprise a \pce{} map with exactly $
2^N $ components equal to $1$s in its diagonal.
For details, we refer the reader to appendix~\ref{quantum_classical}.

%
%


\section{Mathematical considerations} 
\label{sec:math}
This section is devoted to deriving the conditions a Pauli diagonal map needs to satisfy 
the complete positivity condition, i.e., that all the eigenvalues of the Choi matrix
associated to the channel are non-negative.  To do so, we calculate
and diagonalize the Choi matrix of a general Pauli diagonal map, first for a
single qubit and then for $N$ qubits. 
Finally, we restrict from Pauli diagonal maps to PCE maps, and provide a connection
between a vector subspace and the set of coefficients $\{\tau_\valpha\}$ 
in \eref{eq:PCE_definition} of a PCE quantum channel.
This allows us to
derive several important properties of this particular family of channels.
\subsection{Diagonalization of the Choi matrix} 
We now construct the Choi matrix of a single-qubit Pauli diagonal map
$\mathcal{E}$. As described above, $\mathcal{E}$ is a linear map from
$\mcB(\hilbert)$ to itself. We shall denote elements of
$\mcB(\hilbert)$ by the notation $\kket{\cdot}$. Thus, for instance,
$\kket{\sigma_\alpha}$ represents the Pauli matrix $\sigma_\alpha$ understood
as a vector belonging to $\mcB(\hilbert)$, for the present case, in which
$\hilbert={\mathbb{C}^2}$. Since the scalar product in $\mcB(\hilbert)$ is
given by  $\bbrakket{A_1}{A_2}=\tr A_1^\dagger{} A_2$,
elements of the Pauli basis satisfy the relation 
$\bbrakket{{\sigma_{\alpha}} }{{\sigma_{\alpha'} }} = \tr
\left(\sigma_{\alpha}^{\dagger} \sigma_{\alpha'}\right) = 2
\delta_{\alpha\alpha'}$.
In this language, the state of a single qubit reads
$\kket{\rho}=2^{-1}\sum_{\alpha=0}^{3} r_{\alpha} \kket{\sigma_{\alpha}}$
and the matrix form of the map $\mathcal{E}$ is
\begin{equation}
\hat{\mathcal{E}} = \frac{1}{2}\sum_{\alpha=0}^{3} \tau_{\alpha}
\ddyada{\sigma_{\alpha}}.
\end{equation}
After some steps, detailed from \eref{eq:inicio:canal:pauli} to
\eref{Choi_Pauli_1q}, it is possible to show that the Choi matrix of $\mathcal{E}$
reads
\begin{equation}
 \mathcal{D} = \frac{1}{2}\sum_{\alpha=0}^{3} \tau_{\alpha} \sigma_{\alpha}
\otimes \sigma_{\alpha}^*.
\end{equation}
Notice that $\ddyada{\sigma_{\alpha}}$ 
and $\sigma_{\alpha} \otimes \sigma_{\alpha}^*$ are different operators.
Indeed, the former acts as a linear map upon the vector space $\mcB(\hilbert)$,
whereas the latter acts on the tensor product $\hilbert\otimes\hilbert$. Of
course, there is a basis dependent identification between these two
spaces, which is used 
in the construction of the Choi matrix.
%
Surprisingly, one can in fact show that $\mathcal{D}$ is diagonal in the Pauli
basis (see Appendix \ref{sec:appendix:diagonalization}  for details).
%
%
The eigenvalues are
\begin{equation}
 \lambda_{\alpha} = \frac12 \sum_{\beta=0}^3\sa_{\alpha\beta}\tau_{\beta},
\end{equation}
where 
\begin{equation}
\sa=\left(
\begin{array}{cccc}
1 &1 &1   &1   \\
 1 & 1  & -1&-1   \\
1  & -1  & 1 & -1\\
1&-1&-1&1  
\end{array}
\right).
\label{eq:1}
\end{equation}
We wish to add that one can replace $a$ with $H \otimes H$, with $H$ the
Hadamard matrix, and still diagonalize the same Choi matrix $\mathcal{D}$. This is due to the
fact that $a$ corresponds to a permutation of rows of  $H \otimes H$. However,
we chose the aforementioned definition as some later considerations [see
\eref{eq:sigma_property}] cannot be easily written in terms of $H \otimes H$.


%
%

The same program can be carried out for $N$ qubits. In this case, 
one uses the vectorized Pauli strings:
\begin{equation}\label{eq:vectorized_pauli_strings}
\kket{\sigma_{\vec{\alpha} }}=
\kket{\sigma_{\alpha_1}\otimes\cdots\otimes \sigma_{\alpha_N}}.
\end{equation}
This vectorization must not be confused with the 
tensor product of  all $\kket{\sigma_{\alpha_i}}$, 
since the tensor product and the vectorization 
process generally do not commute~\cite{Gilchrist2009}.
%
The vectors satisfy the orthogonality relation
$\bbrakket{{\sigma_{\vec{\alpha} } }}{{\sigma_{\vec{\alpha}'} }} =  2^N
\delta_{\vec{\alpha}\vec{\alpha}'}$. The matrix representation of the
map corresponding to a Pauli diagonal map is 
\begin{equation}
\hat{\mathcal{E}}_N = \frac{1}{2^N}\sum_{\vec{\alpha}} \tau_{\vec{\alpha}} \ddyada{\sigma_{\vec{\alpha} }}.
\end{equation}
As in the previous case, the Choi matrix $\mathcal{D}_N$ may be written in
terms of tensor products of Pauli matrices:
\begin{equation}
 \mathcal{D}_N 
  = \frac{1}{2^N}\sum_{\vec{\alpha}} \tau_{\vec{\alpha}}
       \bigotimes_{j=1}^N \sigma_{\alpha_j} \otimes \sigma_{\alpha_j}^*.
\end{equation}
This matrix is again diagonal in the (multi-qubit) Pauli basis, with the 
eigenvalue corresponding to 
$ \kket{\sigma_{\vec{\alpha} }}$ given by 
%
\begin{equation}
 \lambda_{\valpha} = \frac{1}{2^N} \sum_{\vbeta} \san_{\valpha\vbeta}\tau_{\vbeta},
\label{eq:lambda:is:A:tau}
\end{equation}
where 
\begin{equation}\label{eq:san}
 \san=\sa^{\otimes N}
\end{equation}
Again, the proofs are provided in appendix \ref{sec:appendix:diagonalization}. We wish 
to add that we could diagonalize $\mathcal{D}_N$ with $H^{\otimes 2N}$ instead of
$a^{\otimes N}$, which might be more convenient for other applications.

\subsection{PCE quantum channels as vector spaces} 
\label{sec:vector_spaces}

In this subsection we will provide a one-to-one relation between PCE quantum
channels and the subspaces of a discrete vector subspace associated with the
indices $\valpha$ labeling the components of a state; see
\eref{eq:N_qubits_rho}. 
Some established facts about vector spaces will allow
us to derive the main features of PCE quantum channels.

Let us start by recalling that the problem of determining complete positivity
of a PCE map can be recast as determining which coefficients $\tau_\valpha$ 
are mapped via $\san$
to positive eigenvalues $\lambda_\valpha$, as in \eref{eq:lambda:is:A:tau}. 
Using the fact that $a^{-1}=a/4$, and so 
\begin{equation}
A^{-1} = \frac{1}{4^N} A,
\end{equation}
we can directly invert \eref{eq:lambda:is:A:tau} to obtain 
\begin{equation}
\sum_\vbeta \san_{\valpha \vbeta} \lambda_\vbeta =2^N \tau_\valpha
\label{eq:tau:is:A:p}
\end{equation}
which will serve as a starting point for our analysis. 
This is a remarkable equation, as it provides a method to diagonalize the
Choi matrix of any Pauli diagonal map.

Two other simple but crucial observations are the following. 
For valid quantum channels it holds that 
\begin{equation}
\sum_{\valpha \in \Omega} \lambda_\valpha=0 
\quad \implies 
\quad 
\lambda_\valpha=0, \, \forall \valpha \in \Omega
\label{eq:lambda:is:zero}
\end{equation}
for an arbitrary subset of multi-indices $\Omega$, 
as each member of the sum is greater than or equal to zero, 
due to complete positivity of the underlying channel. 
Finally, setting $\valpha =0$ in \eref{eq:tau:is:A:p}, and taking into 
account the normalization condition that $\tau_{\vec 0}=1$, we obtain
\begin{equation}
\sum_\valpha \lambda_\valpha = 2^N,
\label{eq:sum:alpha:is:2N}
\end{equation}
since, $A_{\vec 0, \vbeta}=1$ for all $\vbeta$. \par
Now we need a definition: to each multi-index $\vec\alpha$ we associate a \textit{set}
of multi-indices $\Phi(\vec\alpha)$ as follows
\begin{equation}
\Phi(\vec\alpha)=\left\{
\vec\beta:\san_{\vec\alpha\vec\beta}=1
\right\}
\label{eq:8}
\end{equation}
If we now assume that $\tau_{\vec\alpha}=1$, and calculate the difference between
\eref{eq:sum:alpha:is:2N} and
$\sum_{\vec\beta}\san_{\vec\alpha\vec\beta}\lambda_{\vec\beta}=2^N$
(which follows from \eref{eq:tau:is:A:p} and $\tau_{\vec\alpha}=1$), one obtains
\begin{eqnarray}
\lambda_{\vec\beta}=0,\quad \forall \vec\beta\notin\Phi(\vec\alpha).
\label{eq:condition:lambda:zero:Phi}
\end{eqnarray}
Thus, if $\tau_{\vec\alpha}=1$, then $\tau_{\vec\gamma}$
and $\tau_{\vec{\gamma^\prime}}$ are equal if
\begin{equation}
\san_{\vbeta\vec\gamma}=\san_{\vbeta\vec{\gamma^\prime}}, \quad
\forall \vec\beta\in\Phi(\vec\alpha).
\label{eq:coneccion:dos:indices:A}
\end{equation}
This follows from restricting the sum \eref{eq:tau:is:A:p}
to the indices $\vec\beta$ such that $\lambda_{\vec\beta}\ne 0$,
given in \eref{eq:condition:lambda:zero:Phi}. 
Condition (\ref{eq:coneccion:dos:indices:A}) therefore connects three
multi-indices, $\valpha$,
$\vec\gamma$ and $\vec{\gamma^\prime}$.
When such a connection exists, $\tau_{\vec\alpha}=1$ implies
$\tau_{\vec\gamma}=\tau_{\vec{\gamma^\prime}}$. 

Let us now work out the nature of the aforementioned connection.  For arbitrary
$k$ we define a vector $\vbeta_k$ such that $\vbeta_k\in\Phi(\vec\alpha)$ as
follows: $\vbeta_k$ is zero everywhere except
for the $k$'th coordinate, which takes a value $\beta$ such that  $\sa_{\alpha_k\beta}=1$. Since 
$\sa_{\alpha 0}=1$ for any $\alpha$, this particular choice of $\vbeta$
indeed belongs to  $\Phi(\vec\alpha)$, so that if \eref{eq:coneccion:dos:indices:A} holds for 
all $\vbeta\in\Phi(\vec\alpha)$, it must hold for that particular $\vbeta_k$, which leads to
%
\begin{equation}
\sa_{\beta\gamma_k}=\sa_{\beta\gamma_k^\prime}
\label{eq:14}
\end{equation}
for all $\beta$ such that $\sa_{{\alpha_k}\beta}=1$.  One can verify, by
working out the different cases, that \eref{eq:14} is equivalently expressed as 
\begin{equation}
\gamma_k^\prime=\alpha_k\oplus{}\gamma_k
\label{eq:oplus}
\end{equation}
where $\oplus{}$ denotes the operation of the Klein group; see Table
\ref{tab:1} for a detailed description. 
\par

It will be useful to think of the multi-index $\valpha$ as an element of 
a vector space. To do so, we notice that any group with the property that
$\alpha \oplus \alpha =0$ is indeed a vector space under the two-element field
$\{0,1\}$.
We notice that the Klein group described in Table \ref{tab:1} is actually isomorphic
to the two-dimensional vector space over the field of two elements $\{0,1\}$.
Then, we build the complete vector space,
with the same field, and defining $\valpha \oplus \vbeta = (\alpha_1 \oplus
\beta_1,\cdots, \alpha_N \oplus \beta_N)$
\footnote{
We can further unravel the vector space, by identifying each index
$\alpha \in \{0,1,2,3\}$ with its binary notation [e.g. identify $2$ with $(1,0)$]
such that to each multindex  $\valpha$ there corresponds a binary string of 
length $2N$. In fact, our sum $\oplus$ would correspond to addition modulo 2 of 
each of the $2N$ components. 
}. 
We can indeed restate \eref{eq:oplus} and say that, for quantum channels, 
if $\tau_{\vec\alpha}=1$, then
$\tau_{\vec\gamma}=\tau_{\vec\alpha\oplus{}\vec\gamma}$.
For example, in \fref{fig:two:qubit:examples}(c) the indices that 
correspond to preserved components are 
$\valpha^{(0)} = (0,0)$, 
$\valpha^{(1)} = (0,2)$, 
$\valpha^{(2)} = (1,0)$, 
$\valpha^{(3)} = (2,2)$, and
$\valpha^{(4)} = (3,2)$. However, 
$\valpha^{(1)} + \valpha^{(2)} = (1,2)$, which is not preserved, and thus this 
diagram does not correspond to a quantum channel.  From this view 
we can derive several interesting observations that will be presented in 
the rest of the section. 

\begin{table} 
\begin{center}
\begin{tabular}{| c | c c c c |}
\hline
$\oplus{}$ & 0 & 1& 2 & 3\\
\hline
0 & 0 & 1 & 2 & 3\\
1 & 1 & 0 & 3 & 2\\
2 & 2 & 3 & 0 & 1 \\
3 &3 & 2 & 1 & 0\\
\hline
\end{tabular}
\caption{Definition of the $\oplus{}$ operation, see \eref{eq:14}. Note that
the operation is an Abelian group; in fact it corresponds to the 
Klein group, where the neutral element is zero. This is the reason for choosing
an additive notation for the operation defined in (\ref{eq:14}). }
\label{tab:1}
\end{center}
\end{table} 

From this readily follows an amusing property: the set of all multi-indices
$\vec\gamma$ for which $\tau_{\vec\gamma}=1$ is closed under binary
vector addition; in other words, it forms a vector subspace of the set
of all multi-indices.
A moment's consideration will further show that the above reasoning can be
inverted; that is, if we set all $\tau_{\vec\gamma}$ equal to $1$
whenever $\vec\gamma$ belongs to a given vector subspace of the set of
all indices, then $\tau$ indeed has an image
which has
only positive components. 
In other words, there is a one-to-one correspondence between 
a quantum channel, and a vector subspace of the aforementioned space. 
\par

With this information, we present a procedure to generate all solutions: we start out
from the solution having $\tau_{\vec0}=1$, with everything else zero. We may
then successively switch $\tau_{\vec\alpha}$'s
to one for various values of $\valpha$,
taking care immediately to set equal to one the components of $\tau$
that
correspond to values of $\vbeta$ generated by the previously switched values of
$\valpha$ via the operation $\oplus$. Doing so, in an ordered way, allows one
to generate
all PCE quantum channels with a given set of preserved components, without the need
of exploring the exponentially large space of all PCE maps.  \par 

\begin{figure} 
\begin{center}
\includegraphics{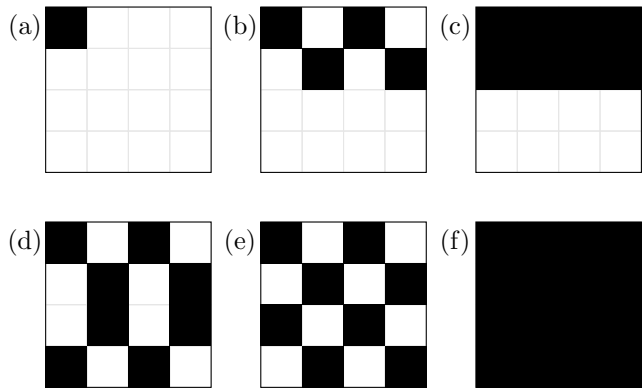}
\end{center}
\caption{
Examples of diagrams for several two-qubit PCE quantum channels: (a) the totally
depolarizing channel and (b) a PCE channel that preserves four components---the
normalization component, one local component of qubit 2, and two correlations between the 
two qubits.
(c), (d), and (e) show the three generators $\mcG_{(1,0)}$, $\mcG_{(3,2)}$, and
$\mcG_{(2,2)}$, respectively; the combination (overlap of diagrams) of any two
of them yields the channel in (b).
(f) represents the identity map. 
}
\label{fig:examplesPCE}
\end{figure} 

%
%

We can show that all PCE quantum channels preserve $2^K$ components. 
First recall that a vector space of dimension $d$ over a field of $q$ elements
has $q^d$ elements~\cite{Roman2008}.
Now $V$ is a vector space on a field of two elements having dimension $2N$. 
We have
seen earlier that
\begin{equation}
W=\left\{
\vec\alpha:\vec\alpha\in V, \tau_{\vec\alpha}=1
\right\}
\label{eq:19}
\end{equation}
is a subspace of $V$. 
As such, $W$ has a given dimension $K$, which means that
$W$ has $2^K$ elements. In other words, a set of indices $\tau_\valpha$ with the property
discussed above can only have $2^K$ elements equal to 1, for a given integer
$K$. \par

It is natural to ask how many PCE quantum channels exist that preserve  $2^K$ components. One
can calculate such number, $\sss_{N,K}$, by examining the number of different
independent subsets of vectors that spawn a given vector subspace. In appendix 
\ref{app:contar} we show that 
\begin{equation}
\sss_{N,K}=\prod_{m=0}^{K-1} \frac{2^{2N-m}-1}{2^{K-m}-1}.
\label{eq:conteo:main}
\end{equation}
From the above expression, it is easy to see the symmetry relation
\begin{equation}\label{eq:conteo:simmetry}
\sss_{N,K}=\sss_{N,2N-K}
\end{equation}
which suggests a relation  between individual channels that
preserve $K$ and $2N-K$ Pauli components that for the time being
has escaped our efforts to identify. 
\par 

Finally, let us point out the following: if we wish to specify a PCE
channel explicitly, the naive way to proceed would be simply to list all the
Pauli components which are not erased. This requires in general, however, an
exponential amount of information: that is, if the system has $N$ qubits, we
generally require of the order of $2^N$ bits to do this. If, on the other hand,
we take advantage of the vector space structure of a PCE channel, we only need
to specify a basis. Since a basis consists of $N$ vectors of length $N$, the
information required is only of $N^2$ bits, so that we have obtained a very
substantial improvement by exploiting complete positivity. This is reminiscent
of a rather similar effect in {\em stabilizer states\/} which can also be
specified by $N^2$ bits, as opposed to an exponentially large number of basis
coefficients for arbitrary states.  A stabilizer state is one which is the
common eigenvector to the eigenvalue 1 of a set of $N$ commuting 
Pauli strings.  The similarity is highly intriguing, and
potentially of interest, since stabilizer states are of central importance in
quantum error correction \cite{Gottesman1997}. 


\section{Generators} 
\label{sec:generators}

We now discuss the existence of a generator set for all PCE quantum channels and 
how to label each of them uniquely as $\pceg$ (according to its local action 
on every qubit in the system). Finally we will discuss a symmetry of
PCE quantum channel generators and a connection between
them and $A$, see \eref{eq:san}.

There exists a subset of PCE quantum channels that generates the entire set; 
the nature of these generators may be studied, as we
shall see, with the properties of the aforementioned vector space.
By standard theorems of linear algebra, any proper subspace $W$, see \Eref{eq:19}, 
can be extended to a maximal non-trivial subspace of dimension $2N-1$ by
adjoining appropriate additional basis elements. This can be done in
different ways. We therefore arrive to the set of maximal extensions of $W$,
where every maximal subspace corresponds to a PCE quantum channel that
preserves half of the Pauli components. The intersection of all the elements of
this set reduces to $W$ itself,
and since intersection of subspaces translates to composition of PCE channels,
this implies that all PCE quantum channels can be obtained as
compositions of PCE channels corresponding to maximally non-trivial subspaces, plus
the identity map. In other words, the set of PCE quantum channels that preserve half 
of the components plus the identity map, is a generator set for 
all PCE channels.
Consider \Fref{fig:examplesPCE}; subfigures (c), (d) and (e) represent nontrivial 
PCE generators (PCEGs) and the composition of any two of them yields the PCE channel
corresponding to (b).

\begin{figure} 
\centering
\includegraphics{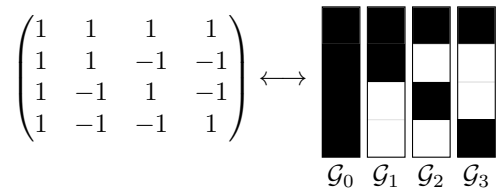}
\caption{
The connection between rows or columns of $a$ and single-qubit PCE
generators $\mathcal G_\alpha$ is shown. One can identify the -1s of $a$ with
0s in the sets $\{\tau_\alpha\}$ of preserved and erased components
of each $\mathcal G_\alpha$.
For any number of particles, such simple relation 
holds, see \sref{sec:generators} and appendix \ref{sec:2_qubits}.
} 
\label{fig:A_and_PCEGs_connection}
\end{figure}

A PCEGs may be characterized by its local
action on every qubit 
in the system. This action can be encoded using a multi-index $\vec\alpha$, 
as in \eqref{eq:pauli_strings}, hence each of the different $4^N$ 
multi-indices may be uniquely related to each of the PCE generators
and thus denoted as $\mcG_{\vec\alpha}$, see figures \ref{fig:A_and_PCEGs_connection} and \ref{fig:appex:twoQ_PCEGs}.
The proof is simplified if one uses the Kraus representation developed in 
\sref{sec:kraus}, so we postpone the demonstration to appendix \ref{sec:2_qubits}.
For single qubits, the identity corresponds to $\mcG_0$, shown in 
\fref{fig:one:qubit:examples}(b), whereas $\mcG_3$ is shown in 
\fref{fig:one:qubit:examples}(c).
The two-qubit PCE generator represented in subfigure (c)
of \Fref{fig:examplesPCE} acts on the first qubit (first column) as a map of its
Bloch sphere to the $x$ axis, and on the second qubit (first row) as an identity,
hence it is labeled $\mcG_{(1,0)}$.
See \Fref{fig:appex:twoQ_PCEGs} for the notation of all two-qubit PCE generators.

A reflection symmetry is identified for PCE generators.
Consider the map $\Sigma^{(k)}$ that reflects a multi-index $\valpha$ with 
respect to the $k$-th axis. This map leaves all components of $\valpha$
invariant, except the $k$-th component, which is transformed according to 
\begin{equation}
0\mapsto3,\quad3\mapsto0,\quad1\mapsto2,\quad2\mapsto1. 
\label{eq:Sigma}
\end{equation}
The maps have the following properties:
\begin{enumerate}
\item $\Sigma^{(k)}(\vec\alpha)\oplus \Sigma^{(k)}(\vec\beta)=\vec\alpha\oplus \vec\beta$, and
\item $\Sigma^{(k)}(\vec\alpha)\neq\vec\alpha$.
\end{enumerate}
From the first property, we now obtain
\begin{equation}
\valpha=\Sigma^{(k)}(\valpha) \oplus \Sigma^{(k)}(\vec0). 
\label{eq:Sigma:oplus}
\end{equation}
This implies that if $\Sigma^{(k)}(\vec0)$ belongs
to a channel then 
$\tau_\valpha = \tau_{\Sigma^{(k)}(\vec\alpha)}$, where we used 
the fact that for channels the non-zero elements are closed
under $\oplus$. In other words, the components of a PCE channel
are symmetric under reflection over the $k$th axis. Now consider the case in which 
 $\Sigma^{(k)}(\vec0)$ does not belong to generator. Then 
$\tau_\valpha \ne \tau_{\Sigma^{(k)}(\vec\alpha)}$, since
the case $\tau_\valpha = \tau_{\Sigma^{(k)}(\vec\alpha)}=1$
is forbidden due to \eref{eq:Sigma} and the case 
$\tau_\valpha = \tau_{\Sigma^{(k)}(\vec\alpha)}=0$ is also forbidden 
because for generators the codimension of the associated vector space 
is 1. This means that the components of a PCE channel are antisymmetric
with respect to reflection over the $k$th axis. 
Indeed, the two-qubit PCE generators 
$\mcG_{(1,0)}$, $\mcG_{(3,2)}$ and $\mcG_{(2,2)}$
represented in Figs.~\ref{fig:two:qubit:examples}(c),
\ref{fig:two:qubit:examples}(d), and \ref{fig:two:qubit:examples}(e),
respectively, are either symmetric or anti-symmetric under reflection with 
respect to lines that divide the diagram in half vertically and horizontally. 

%
%
%

Finally, it is worth pointing out that $\san$ (and thus $a$) [see
\eref{eq:san}] encodes all the information of PCE generators $\pceg$ and,
therefore, of all PCE quantum channels.  
%
%
%
%
From $A$, the tensor power of matrix $a$, one can infer the components
$\{\taus\}$ of a PCE generator $\pceg$ by taking
row (or column) $\valpha$ of $A$ and replacing $-1$ with
$0$.
The proof of the connection between PCEGs and $A$ is
given in \appref{sec:2_qubits}, and in
\fref{fig:A_and_PCEGs_connection} we illustrate this connection for the
single-qubit case.

%


\section{PCE channels and decoherence} 
\label{sec:kraus}

Lindblad processes arise naturally in many
theoretical~\cite{lindblad,kossa,kossa2,Gorini1976,cirac,davalosdivisibility}
and experimental~\cite{PhysRevA.67.042322} settings and are archetypical in
decoherence dynamics. Moreover, these processes lead to a monotonic
(continuous) loss of information~\cite{breuerreview} and describe
non-invertible channels in the asymptotic limit $t\to \infty$ (this can be seen
from the monotonic (continuous) decrease of the determinant, see
Ref.~\cite{cirac}).
It is known that not every quantum channel can be seen as a snapshot of a
process
arising from a traditional Lindblad equation or even a time-dependent Lindblad
equation~\cite{cirac,davalosdivisibility}. 
Therefore, an interesting question is whether \pce{} channels can be seen as limit points of
some Markovian processes. In this section we prove that in fact they are, and
give two examples of Markovian implementations.
The first of them consists in identifying each \pce{} channel as a fixed point
of some pure dissipative process, and in the second implementation we relate
each \pce{} channel to fixed
points of some memory-less collision model.
\subsection{Kraus representation} 
To derive the aforementioned implementations, we exploit the existence of the
PCEGs and their \textit{Kraus representation} (or operator-sum representation)
which, for an arbitrary channel $\mcE$,  reads 
\begin{equation}
\mcE[\rho]=\sum_i K_i \rho
K_i^\dagger{},
\end{equation}
with $\sum_i K_i^\dagger{}K_i=\one$ (the trace-preserving condition)~\cite{Kraus1983}.
Inspection of the Kraus operators for two-qubit PCEGs leads to
the ansatz that the Kraus operators for the generator of 
$\mcG_{\vec \alpha}$ are 
\begin{equation}
K_0=\frac{\openone}{\sqrt2}, \quad K_1=\frac{\sigma_{\vec \alpha}}{\sqrt2},
\label{eq:kraus:PCE}
\end{equation}
since the Kraus operators corresponding to a single-qubit PCE are 
$\{\openone/\sqrt2, \sigma_\alpha/\sqrt2\}$,
corresponding to the operation that leaves the component $\sigma_\alpha$
invariant~\cite{nielsen_chuang_2011}. Notice that according to Kraus operators
the generators $\mcG_{\vec \alpha}$ are $N$-qubit flip channels with flip
probability $1/2$, where the joint flip is $\rho \mapsto \sigma_{\vec
\alpha} \rho \sigma_{\vec \alpha}$. In fact, tracing out all particles except
the $k$th one gives the well-known qubit flip channels, \ie{}, $\tr_{\not k}
\mcG_\valpha=\mcG_{\alpha_k}$, see~ \eref{eq:generator:individual:qubit}. 
More generally, tracing out $m$ particles leaves a $N-m$ particles flip channel
(completely dephasing).

We shall first show that the Kraus operators in \eref{eq:kraus:PCE} produce a PCE. 
Notice that 
$\sigma_\alpha \sigma_\beta \sigma_\alpha =a_{\alpha \beta} \sigma_\beta$, see~\Eref{eq:1},
which in turn implies that 
\begin{equation}
\sigma_{\vec \alpha} \sigma_{\vec \beta} \sigma_{\vec \alpha} 
=\san_{\vec \alpha \vec \beta} \sigma_{\vec \beta}.
\label{eq:sigma_property}
\end{equation}
Next, consider the action of a channel with Kraus representation 
\eqref{eq:kraus:PCE} on a $N$-qubit system:
\begin{align}
\rho \mapsto &\frac{1}{2^N}\sum_{\vec \beta} r_{\vec \beta} \left( \frac{1}{2} \sigma_{\vec \beta} + \frac{1}{2} \sigma_{\vec \alpha} \sigma_{\vec \beta} \sigma_{\vec \alpha} \right)\nonumber \\
=&\frac{1}{2^N}\sum_{\vec \beta} r_{\vec \beta} \frac{1+A_{\vec \alpha \vec \beta}}{ 2}\sigma_{\vec \beta}.
\end{align}
However, since $\san_{\vec \alpha \vec \beta} = \pm 1$, the channel characterized
by the Kraus operators in \eref{eq:kraus:PCE} is a PCE channel. Moreover, one can notice that, 
except for the first row, half of the matrix elements of each row are $+1$ and 
half are $-1$, which implies that the channel is a PCEG. 


Observe	 also that a different choice of $\vec \alpha$ in \eref{eq:kraus:PCE}
leads to different channels. This follows from the fact that if two channels were
the same this would imply that the matrix representation of the corresponding
superoperator of $\rho \mapsto \sigma_{\vec \alpha}\rho \sigma_{\vec \alpha}$
would have to be the same, which is clearly false. Since there are $4^n$ different 
$\vec \alpha$ values, this implies that all PCEGs (plus the identity map) are in 
one-to-one correspondence.  

\subsection{Pure dissipative implementation} 
In this section we show that any PCE channel can be seen as the
fixed point of some decoherence process, starting with PCEGs and 
then extending to more general channels. 
Consider the following dynamical process that
implements $\mcG_{\vec \alpha}$ when $t\to \infty$,
\begin{align}
\mcG_{\vec \alpha,t}[\rho]&=e^{- \gamma t} \rho + (1-e^{-\gamma t})\mcG_{\vec \alpha}[\rho]\nonumber \\
&=\frac{\left( 1+e^{- \gamma t} \right)}{2} \rho +\frac{\left( 1-e^{- \gamma t} \right)}{2} \sigma_{\vec \alpha} \rho \sigma_{\vec \alpha},
\end{align}
where $\gamma>0$. It is easy to show that the family of channels $\mcG_{\vec
\alpha, t}$ parametrized with $t \geq 0$ forms a one-parametric semigroup,
\ie{} $\mcG_{\vec \alpha, t_1}\mcG_{\vec \alpha, t_2}=\mcG_{\vec \alpha,
t_1+t_2}$. Therefore $\mcG_{\vec \alpha, t}$ describes a dissipative
time-homogeneous Markovian process, which is always characterized by some
Lindblad generator~\cite{lindblad}. The Lindblad generator of $\mcG_{\vec
\alpha, t}$, denoted by $\mcL_{\vec \alpha}$, can be obtained using the standard
procedure:
\begin{equation}
\mcL_{\vec \alpha}[\rho]
   =\left. \frac{d\mcG_{\vec \alpha,t}[\rho]}{dt}\right|_{t=0}
   =\frac{\gamma\left(\sigma_{\vec \alpha}\rho \sigma_{\vec \alpha} -\rho \right)}{2},
\end{equation}
where the unique Lindblad operator associated with the relaxation ratio
$\gamma/2$ is simply $\sigma_{\vec \alpha}$. Notice that $\sigma_{\vec \alpha}$
is trace-less, therefore the process is purely dissipative~\cite{cirac}.

Since PCEGs commute, we can describe easily any other PCE channel as a fixed
point of a decoherence process. For them, the Lindblad generators are the sum
of the Lindbladians of the corresponding generators. As an example, consider
the channel depicted in \fref{fig:two:qubit:examples}(b); it is equal to
$\mcG_{(0,3)}\mcG_{(3,3)}$, therefore it is the fixed point of the dissipation
process described with the following Lindbladian:
\begin{equation}
\mcL[\rho]=\frac{\gamma_{(0,3)}\left(\sigma_{(0,3)}\rho\sigma_{(0,3)}-\rho \right)}{2}
   +\frac{\gamma_{(3,3)}\left(\sigma_{(3,3)}\rho\sigma_{(3,3)}-\rho
\right)}{2},
\end{equation}
where $\gamma_{(0,3)}$ and $\gamma_{(3,3)}$ are positive and correspond to the Lindblad operators $\sigma_{(0,3)}$ and $\sigma_{(3,3)}$. Notice that such election of Lindblad operators is not unique, as the PCE channel described here is also equal to $\mcG_{(0,3)}\mcG_{(3,0)}$.

\subsection{Collision model implementation} 
We show now that PCE channels can also be implemented with \textit{simple
collision models}~\cite{Ziman2010}. To do this, observe that employing the
Stinespring dilation theorem~\cite{Stinespring2006} PCEGs can
be implemented using a unitary over the system and an
ancilla. Since PCEGs
always have Kraus rank $2$, one can
always choose a qubit as the ancillary system. Concretely,
\begin{equation}
\mcG_{\vec \alpha}[\rho]=
    \tr_{\text{qubit}} \left( 
         U_{\vec \alpha} \left(\rho\otimes \projj{0}\right) U_{\vec \alpha}^\dagger{} 
     \right),
\end{equation}
where $\tr_{\text{qubit}}$ denotes the partial trace over the ancillary qubit, with
the unitary defined as follows:
\begin{align}
U_{\vec\alpha}\ket{\psi} \ket{0}= \frac{1}{\sqrt{2}} 
   \left( \ket{\psi} \ket{0}+\sigma_{\vec\alpha} \ket{\psi}\ket{1} \right),\\
U_{\vec\alpha}\ket{\psi} \ket{1}=\frac{1}{\sqrt{2}} 
   \left( \ket{\psi} \ket{0}-\sigma_{\vec\alpha} \ket{\psi}\ket{1} \right).
\label{eq:Nqubit_stinespring}
\end{align}
Therefore, any concatenation of PCEGs can be described as a collision model
with as many collisions as generators needed. In fact, generators are described
with one collision. For the general case consider some PCE channel $\mcE$
generated with $\left\{ \mcG_{\vec \alpha_1}, \mcG_{\vec \alpha_2}, \dots,
\mcG_{\vec \alpha_M} \right\}$. For this we can define an environment
consisting of $M$ qubits initially in the state $\left( \projj{0}
\right)^{\otimes M}$, or equivalently one qubit with the additional assumption
that its state is reset to $\ket{0}$ after each collision (memory-less
collisions). The collision with the $k$-\text{th} particle is described by
$U_{\vec \alpha_k}$, which acts solely over the system and the $k\text{th}$
particle. Therefore $\mcE$ can be written as follows,
\begin{equation}
\mcE[\rho]=\tr_{\text{E}} \left[ \left(U_{\vec \alpha_1} \dots U_{\vec \alpha_M}\right) \rho \otimes \left(\projj{0}\right)^{\otimes M} \left(U_{\vec \alpha_1} \dots U_{\vec \alpha_M} \right)^\dagger{} \right],
\end{equation}
where $\tr_{\text{E}}$ is the partial trace over all ancillary qubits. Notice
that as PCEGs commute the order of the collisions is irrelevant.
%


\section{Conclusions and outlook}
\label{sec:conclusions}

In this paper we introduce and characterize a set of quantum maps which either preserve
or completely erase the components of a multi-qubit density matrix, in the
basis of Pauli strings; 
we call  those maps {\it Pauli component erasing  maps}.
For a single qubit these include the completely depolarizing
and dephasing channels.
To start the characterization, we note that 
not all PCE maps are quantum channels, as some are not completely positive.  In
fact, the most laborious task of this paper was to evaluate complete positivity conditions
given by the Choi-\jami{} isomorphism, after which we showed that
the components of PCE quantum channels form a finite vector space.
This in turn allows us to unravel several properties, such as the possible number of PCE
channels, and the number of components preserved while also providing 
advantages to study numerically this set, for example by implying an efficient method 
to construct all quantum channels for a given number of qubits.
%

Similar to other objects in open quantum systems (for example, Lindblad
processes), PCE quantum channels form a semigroup, but finite in this case.
For
PCE channels, the generators  are generalized flip operations, \ie{} channels
that with probability $1/2$ apply a joint flip. 
This structure allows us to link this channel with multi-qubit decoherence
processes which can be described, say, by simple dissipative processes or
memory-less collision models, which in turn may pave a way to either implement
these channels or connect them with already existing decoherence families. 
This, together with the discovered algebraic structure that translates complete
positivity into an explicit conditioned preservation of many-body correlations,
encompasses an advance in the knowledge of the mathematical structures
underlying general quantum channels.


In the future we might consider generalizations (such as going from qubits to
qudits) as well as the geometric role of PCE channels within the set of all
quantum channels to further advance the
understanding of open quantum systems.  We have thus described a family of
quantum channels with a very special mathematical structure that allows us to
widen the understanding of quantum channels in the context of many-body
systems.

\begin{acknowledgments}
Support by CONACyT Grants No. 285754 and No. 254515 and UNAM-PAPIIT Grants No. IG100518 and 
No. IG101421 is
acknowledged.  
J. A. d.~L. acknowledges a scholarship from CONACyT.
J. A. d.~L. thanks Juan Diego Chang for the fruitful
discussions and support at the early stages of this project.  A. F.
acknowledges funding by Fundação de Amparo à Ciência e Tecnologia do Estado de
Pernambuco, through Grants No. BFP-0168-1.05/19 and No. BFP-0115-1.05/21.
D. D. acknowledges OPTIQUTE Grant No. APVV-18-0518, DESCOM Grant No. VEGA-2/0183/21, and the Štefan
Schwarz Support Fund.
\end{acknowledgments} 
\appendix
\section{Quantum-Classical channels}
\label{quantum_classical}
A quantum-classical (QC) channel is defined by using an orthonormal basis in
the Hilbert space. Let $B=\left\{ \ket{\psi_i} \right\}$ be such a basis in
$\hilbert$ with $\dim(\hilbert)=2^N$; the QC channel associated to $B$ is
\begin{equation}
\mcE^{\text{QC}}_B[\rho]=\sum_{i=1}^{2^N} \bra{\psi_i}\rho \ket{\psi_i} \projj{\psi_i},
\end{equation}
that is, QC channels project density matrices onto the corresponding diagonal
matrix in the basis $B$~\cite{nathanson2007pauli}.

Consider now that the basis $B$ is the simultaneous eigenbasis of a maximal set
of commuting Pauli strings denoted by $\text{set}(B)$; such a set contains $2^N$
elements (including the identity), and there are $2^N+1$ of such
sets~\cite{Lawrence2002}. Now we proceed to demonstrate that the QCs defined in
this way are \pce{} channels with $2^N$ $1$s on their diagonal.

First we compute the components of $\mcE^{\text{QC}}_B$ in the basis
$2^{-N/2}\left\{\sigma_{\vec \alpha}\right\}$:
\begin{equation}
\left(\mcE^{\text{QC}}_B\right)_{\vec k \vec l}=\frac{1}{2^N}\sum_{i=1}^{2^N} \bra{\psi_i}\sigma_{\vec k} \ket{\psi_i}\bra{\psi_i}\sigma_{\vec l} \ket{\psi_i}.
\end{equation}
To evaluate the components, observe that $\bra{\psi_i}\sigma_{\vec k}
\ket{\psi_i}^2=1 \ \ \forall \sigma_{\vec k}\in \text{set}(B)$, and from the
formula for the purity of $\ket{\psi_i}$,
\begin{equation}
1=\frac{1}{2^N}\sum_{\vec k} \bra{\psi_i} \sigma_{\vec k} \ket{\psi_i}^2,
\end{equation}
it follows that $\bra{\psi_i} \sigma_{\vec k'} \ket{\psi_i}=0 \ \ \forall
\sigma_{\vec k'} \not\in \text{set}(B)$ since there are only positive terms in
the sum, and $2^N$ of them are already equal to $1$. Therefore,
\begin{align}
\left(\mcE^{\text{QC}}_B\right)_{\vec k \vec k}&=1,\nonumber\\ \left(\mcE^{\text{QC}}_B\right)_{\vec k \vec l'}&=\left(\mcE^{\text{QC}}_B\right)_{\vec l' \vec l'}=0 \ \ \forall \sigma_{\vec k} \in \text{set}(B) \ \ \forall \sigma_{\vec l'} \not\in \text{set}(B).
\end{align}
To compute $\left(\mcE^{\text{QC}}_B\right)_{\vec k \vec l}$ for $\sigma_{\vec k}, \sigma_{\vec l} \in \text{set}(B)$ with $\vec k \neq \vec l$, observe that $\left(\mcE^{\text{QC}}_B\right)_{\vec k \vec l}=\left(\mcE^{\text{QC}}_B\right)_{\vec l \vec k}$, \ie{} the matrix corresponding to $\mcE^{\text{QC}}$ is an orthogonal projector. Thus, considering the block,
\begin{equation}
\left[ \begin{array}{cc}
\left(\mcE^{\text{QC}}_B\right)_{\vec k \vec k} & \left(\mcE^{\text{QC}}_B\right)_{\vec k \vec l} \\ 
\left(\mcE^{\text{QC}}_B\right)_{\vec k \vec l} & \left(\mcE^{\text{QC}}_B\right)_{\vec l \vec l}
\end{array}  \right]=\left[ \begin{array}{cc}
1 & \left(\mcE^{\text{QC}}_B\right)_{\vec k \vec l} \\ 
\left(\mcE^{\text{QC}}_B\right)_{\vec k \vec l} & 1
\end{array}  \right],
\end{equation}
it is easy to check that the latter is a projector only if $\left(\mcE^{\text{QC}}_B\right)_{\vec k \vec l}=0$. Since there are exactly $2^N$ elements of the form $\left(\mcE^{\text{QC}}_B\right)_{\vec k \vec k}$ with $\sigma_{\vec k} \in \text{set}(B)$, then the channel $\mcE^{\text{QC}}_B$ is \pce{} with $2^N$ $1$s on its diagonal.

\section{Diagonalization of Choi Matrix $\mathcal{D}_N$} 
\label{sec:appendix:diagonalization}
In order to simplify the derivation of the relations, let us employ pairs of
binary indices instead of a single quaternary, i.e., $\alpha\to j +2 k$. For
the sake of clarity, we use Latin symbols for binary indices, and
reserve Greek letters for quaternary ones. 
We can write the elements of the Pauli basis
$(\sigma_{0},\sigma_{1},i\sigma_{2},\sigma_{3})$, 
compactly as
$\sigma_{k l }=\sum_{j=0}^{1} (-1)^{j k }\dyad{j}{j + l ( \text{mod}2) }$. In vectorized form
\begin{equation}
\kket{\sigma_{k l }}=\sum_{j=0}^{1} (-1)^{j k }\ket{j} \ket{j+l(\text{mod } 2) },
\label{CompToPau}
\end{equation}
and its inverse relation
\begin{equation}
\ket{k}\ket{k+l (\text{mod } 2)}=\frac{1}{2}\sum_{j=0}^{1} (-1)^{jk}\kket{\sigma_{jl}}.
\label{PauToComp}
\end{equation}
On the other hand, the matrix form of an arbitrary Pauli map $\mathcal{E}$ may
be written as
\begin{align}
\hat{\mathcal{E}} & = 
\frac{1}{2}\sum_{l m=0}^{1} \tau_{l m} \ddyada{\sigma_{l m}} 
\label{eq:inicio:canal:pauli}\\
& = \frac{1}{2}\sum_{jkl m} \tau_{l m} (-1)^{l (j+k)} \nonumber\\
     & \hspace{1cm} \times \ket{j}\ket{j +m  (\text{mod } 2)}\bra{k}\bra{k +m (\text{mod } 2)}.
\end{align}
After applying the reshuffling operation on $\hat{\mathcal{E}}$, we obtain the
Choi matrix associated to the map. It reads
\begin{multline}
 \mathcal{D} = 
  \frac{1}{2}\sum_{jklm} \tau_{lm} (-1)^{l(j+k)} \\
     \times \ket{j}\ket{k}\bra{j+ m(\text{mod}2)}\bra{k+m(\text{mod}2)}.
\label{Choi_a1}
\end{multline}
Note furthermore that the expression above may also be written as a combination
of tensor products of Pauli matrices
\begin{equation}
 \mathcal{D} = 
   \frac{1}{2}\sum_{l m } \tau_{l m } \sigma_{l m } \otimes \sigma_{l m }^*.
 \label{Choi_Pauli_1q}
\end{equation}
Returning to the Eq. \ref{Choi_a1}), let us apply the index relabeling $k\to j+
k (\text{mod} 2)$; then the Choi matrix reads
\begin{multline}
 \mathcal{D} =
    \frac{1}{2}\sum_{jk lm} \tau_{ lm} (-1)^{ l k} \\
       \times \ket{j}\ket{j +k(\text{mod } 2)}\bra{j+ m(\text{mod } 2)}\bra{j+ m+ k(\text{mod } 2)},
\nonumber
\end{multline}
since $ (-1)^{j+(j+ k (\text{mod} 2))} = (-1)^k $. To continue, we 
use the relation between computational and Pauli elements [\eref{PauToComp}],
and notice that $\sum_j(-1)^{j(m\pm n)}=2\delta_{mn}$. We arrive to the
simple expression 
\begin{align}
 \mathcal{D} & = 
  \frac{1}{2}\sum_{jk} 
      \left(\frac{1}{2}\sum_{l  m } 
(-1)^{j m + k l}
\tau_{l  m } 
\right) 
     \ddyad{\sigma_{jk}}{\sigma_{jk}}.
\label{eq:D:diagonalize}
\end{align}
Notice that  $\mathcal{D}$ is already written in its diagonal form, and one can 
identify by inspection the eigenvalues. 
The eigenvalues read 
\begin{equation}
\lambda_{jk}=\frac{1}{2}\sum_{l m } (-1)^{j m + k l} \tau_{l m },
\label{eq:lambda:is:tau:appendix}
\end{equation}
or more compactly ${\lambda} =(1/2) H \otimes H {\tau}$, where 
$H$ is the Hadamard matrix.

For the sake of convenience in the demonstration of several useful properties
of the PCE channels, 
we shall reorder the eigenvalues, to write 
\begin{equation}
\lambda =  \frac12 \sa \tau
\end{equation}
with $a$ the matrix shown in~\eref{eq:1} 
instead of $H\otimes H$. This can be done due to the fact
that both matrices ($\sa$ and $H \otimes H$) are equivalent up to a permutation of rows. 
In other words this operation corresponds to a reordering of the eigenvalues.

\subsection{$N$ qubits} 
To work out the $N$-qubit case, we again rely on binary indices. In this case, we replace 
$N$-dimensional vector $\valpha$ with a pair of $N$-dimensional vector binary
indices $\vec j$ and $\vec k$ so that 
each entry $\alpha_i$ of $\valpha$ is identified with the pair $j_i$ and $k_i$ as in the 
single-qubit case of the previous subsection. Then, all the steps leading to 
\eref{eq:lambda:is:tau:appendix} can be redone.

The tensor product of Pauli matrices, in vector form,  will be denoted by 
$\kket{\sigma_{\vec{k}\vec{l} }}$.
With this in mind, a $N$-qubit Pauli map can be written as 
%
%
\begin{equation}
\hat{\mathcal{E}}_N = 
  \frac{1}{2^N}\sum_{\vec{l}\vec{m}} \tau_{\vec{l}\vec{m}} \ddyada{\sigma_{\vec{l}\vec{m} }}. 
\end{equation}
The generalizations of Eqs.~(\ref{CompToPau}) and (\ref{PauToComp}) read 
\begin{align}
\kket{\sigma_{\vec{k}\vec{l} }}  & = 
   \sum_{\vec{j}} (-1)^{\vec{j}\cdot\vec{k}} \ket{\vec{j}}\ket{\vec{j}+ \vec{l}(\text{mod}2)},\\
 \ket{\vec{l}}\ket{\vec{l}+ \vec{n}(\text{mod}2)} & = 
   \frac{1}{2^N}\sum_{\vec{m}} (-1)^{\vec{m}\cdot\vec{l}}\kket{\sigma_{\vec{m}\vec{n} }}.
\end{align}

%
By employing the previous relations, we can write the matrix representation of
the map, $\hat{\mathcal{E}}_N $ in the $N$-qubit computational basis as
\begin{multline}
\hat{\mathcal{E}} 
= \frac{1}{2}\sum_{\vec j\vec k\vec l \vec m} 
     \tau_{\vec l \vec m} (-1)^{\vec l (\vec j+\vec k)} \\
     \times\ket{\vec j}\ket{\vec j+\vec m (\text{mod}2)}\bra{\vec k}\bra{\vec k+\vec m(\text{mod}2)}.
\end{multline}
%
%
In this way it is straightforward to apply the reshuffling operation on
$\hat{\mathcal{E}}_N$ to obtain the associated Choi matrix, and
then transform back to the Pauli basis and simplify to obtain 
\begin{align}
\mathcal{D}_N & = 
   \frac{1}{2^{N}}\sum_{\vec{m}\vec{n}} \left( 
        \frac{1}{2^{N}}\sum_{\vec{l}\vec{m}} \tau_{\vec{l}\vec{m}} (-1)^{\vec{l}\cdot\vec{n}+\vec{m}\cdot\vec{m}} \right)\ddyad{\sigma_{\vec{m}\vec{n} }}{\sigma_{\vec{m}\vec{n} }}.
\end{align}
All intermediate steps, from \eref{eq:inicio:canal:pauli} to 
\eref{eq:D:diagonalize} are similar, but with a vectorized version of the 
indices, and appropriate normalization constants. 
Again, we are left with an expression that displays explicitly the eigenvalues of 
the Choi matrix, so we can write 
\begin{equation}
\lambda_{\vec{j}\vec{k}}= \frac{1}{2^{N}}\sum_{\vec{l}\vec{m}} 
    (-1)^{\vec{j}\cdot\vec{m}+\vec{k}\cdot\vec{l}}
    \tau_{\vec{l}\vec{m}} 
\end{equation}
or more compactly $\lambda = (H\otimes H/2)^{\otimes N} \tau$. 
Again, we prefer to reorganize the indices to be able to write
\begin{equation}
\lambda = \frac{1}{2^N} A \tau, 
\end{equation}
where 
$\san=a^{\otimes N}$. 


\section{Number of PCEs for a fixed number of invariant components} 
\label{app:contar}
Finally, we may enumerate straightforwardly the subspaces $W$ of dimension $K$.
We do this in two steps: first, we evaluate $\nnn_{K,N}$, the number of all
linearly independent subsets $V$ with $K$ elements. Each of these is the basis
of one subspace of dimension $K$, but each subspace has a number $\mmm_K$ of
different bases. The crucial point is that $\mmm_K$ is independent of the
subspace under consideration: $\mmm_K$ simply describes the number of linear
maps of $W$ onto itself. The total number $\sss_{N,K}$ of subspaces of
dimension $K$ is therefore $\nnn_{N,K}/\mmm_K$. 

To evaluate $\nnn_{N,K}$ we proceed by steps: the first element of the basis
can be any non-zero element, of which the number is $2^{2N}-1$. For the  basis
element $m+1$, we must choose from those which do not belong to the $m$
dimensional space generated by the first $m$ basis elements, so that one
chooses from $2^{2N}-2^{m}$. We thus have
\begin{equation}
\nnn_{N,K}=\prod_{m=0}^{K-1}\left( 2^{2N}-2^{m} \right).
\label{eq.20}
\end{equation}
On the other hand, any map of a $K$-dimensional vector space $W$ onto itself is
uniquely defined by a non-singular binary $K\times K$ matrix over the field
$\{0,1\}$. To count these, we proceed as above: the first line is an arbitrary
non-zero vector, of which there are $2^K-1$. For the row $m+1$  we must choose
an arbitrary vector not belonging to those generated by the first $m$ vectors,
of which there are $2^K-2^{m}$.  This eventually yields
\begin{equation}
\mmm_{K}=\prod_{m=0}^{K-1}\left(
2^{K}-2^{m}
\right).
\label{eq.21}
\end{equation}
From this it follows that
\begin{equation}
\sss_{N,K}=\prod_{m=0}^{K-1} \frac{2^{2N-m}-1}{2^{K-m}-1}.
\label{eq:22}
\end{equation}
\section{Local action and labeling of PCE generators}
\label{sec:2_qubits}

\begin{figure} 
\centering
\includegraphics[width=\columnwidth]{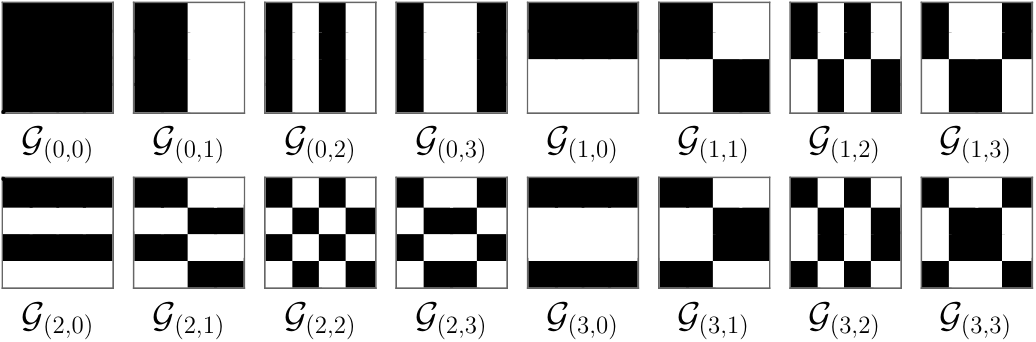}
\caption{PCE generators for two qubits. Notice that all generators
are either symmetric or anti-symmetric under
horizontal and vertical reflections. }
\label{fig:appex:twoQ_PCEGs}
\end{figure} 

The local action of a generator $\mcG_\valpha$ on every qubit in the 
system depends only, as its notation suggests, on the multi-index $\valpha$. 
This index has a simple meaning that can be read from the graphical representation
of the channel. 
Recall the single qubit PCE generators, shown in 
\fref{fig:A_and_PCEGs_connection}, denoted by  $\mcG_0$ (corresponding to the identity map) 
and $\mcG_{1,2,3}$ (corresponding to the completely bit, phase and bit-phase
flip channels respectively).
One can easily read the diagrams in the following 
manner: $\alpha =0$ corresponds to all squares black, whereas for $\alpha>0$ we have
only the zeroth and the $\alpha$-th squares black. 
Let us  generalize this characterization rule for $N$-qubit 
PCE generators. 
Consider that the reduced 
density matrix of the $k$th qubit after generator $\mcG_\valpha$ acts
on the entire system
\begin{align}
\tr_{\not k} \mcG_\valpha [\rho] &= 
\frac12 \tr_{\not k} (\rho + \sigma_\valpha \rho \sigma_\valpha)\nonumber \\
&= \frac{\rho_k}{2} + \frac{\sigma_{\alpha_k} \rho_k \sigma_{\alpha_k}}{2}\nonumber \\
&= \mcG_{\alpha_k} [\rho_k ],
\label{eq:generator:individual:qubit}
\end{align}
where $\not k$ means that all qubits except for the $k$th one are traced out.
We can read from \eqref{eq:generator:individual:qubit} that 
$\alpha_k$ not only characterizes $\mcG_{\alpha_k}$ but actually tells 
us which single qubit channel is acting locally on the $k$th qubit.  The
action of $\mcG_{\alpha_k}$ on the local components of the reduced density
matrix $\rho_k$ reads 
$r_{0,\ldots,j_k,\ldots,0}\mapsto 
\tau_{0,\ldots,j_k,\ldots,0}r_{0,\ldots,j_k,\ldots,0}$.
%
The general characterization rule for all PCE generators $\mcG_\valpha$ is clear
now: if all $\tau_{0,\ldots,j_k,\ldots,0}=1$, then $\alpha_k=0$;
otherwise, 
if $\tau_{0,\ldots,j_k,\ldots,0}=1$ (with $j_k>0$), then 
$\alpha_k=j_k$.
%
%
For two-qubit PCE diagrams this means
that the multi-index $\valpha$ is encoded in the first column and row of 
the diagrams. 
For example, see $\mcG_{(0,2)}$ in
     \fref{fig:appex:twoQ_PCEGs}), where all
     $\tau_{j_1,0}=1$ and $\tau_{(0,2)}=1$, and thus $\valpha=(0,2)$.
%
In \fref{fig:appex:twoQ_PCEGs} we show all two-qubit PCE generators
and their corresponding notation $\mcG_\valpha$.

An interesting relation of the generators and the $A$ matrix can be derived 
with the tools developed. 
Consider the generator 
$\mcG_\valpha$, and its Pauli components $\tau^{(\valpha)}_\vbeta$. We 
can calculate the former studying the action of the generator on the 
non-normalized state $\varrho =  \sum_\vgamma \sigma_\vgamma$.
Let us proceed with such calculation, using the Kraus decomposition \eref{eq:kraus:PCE}:
\begin{align}
\tau^{(\valpha)}_\vbeta &= \tr \sigma_\vbeta \mcG_\valpha[\varrho] \\
&= \frac12 \sum_\vgamma \tr \left[ \sigma_\vbeta \sigma_\vgamma
+ \sigma_\vbeta \sigma_\valpha \sigma_\vgamma \sigma_\valpha \right] \\
&= \frac12 \left(1+ A_{\valpha \vbeta} \right)
\end{align}
where we have used the orthogonality relations of Pauli matrices and \eref{eq:sigma_property}.
This means that one can read the $\alpha$-th generators directly from matrix $A$, see 
\fref{fig:A_and_PCEGs_connection} for the $n=1$ case. Alternatively 
one could construct the $A$ matrix for $n=2$, from \fref{fig:appex:twoQ_PCEGs}, 
where the first row of this matrix is read from $\mcG_{(0,0)}$, replacing
black (white) squares with 1's ($-1$s), the second row from $\mcG_{(0,1)}$, etc.

\bibliography{references}
\end{document}